\documentclass[reprint,superscriptaddress, amsmath,amssymb, floatfix, pra]{revtex4-2}

\usepackage{xcolor}
\usepackage{graphicx}
\usepackage{braket}
\usepackage{mathtools}
\usepackage[colorlinks, citecolor=blue]{hyperref}
\usepackage[utf8]{inputenc}
\usepackage[T1, OT1]{fontenc}
\usepackage{float}
\usepackage[caption = false]{subfig}
\usepackage{dsfont}
\usepackage{pgfplots}

\DeclarePairedDelimiter\floor{\lfloor}{\rfloor}

\newcommand{\tr}{\textcolor{red}}

\begin{document}
\title{Quantum direct communication protocol using recurrence in k-cycle quantum walk}
\author{Sanjeet Swaroop Panda}
\email{sanjeetpanda@iisc.ac.in}
\author{P. A. Ameen Yasir}
\email{ameenyasir.p.a@gmail.com}
\affiliation{Quantum Optics \& Quantum Information,  Department of Instrumentation and Applied Physics, Indian Institute of Science, Bengaluru 560012, India}
\author{C. M. Chandrashekar}
\email{chandracm@iisc.ac.in}
\affiliation{Quantum Optics \& Quantum Information,  Department of Instrumentation and Applied Physics, Indian Institute of Science, Bengaluru 560012, India}
\affiliation{The Institute of Mathematical Sciences, C. I. T. Campus, Taramani, Chennai 600113, India}
\affiliation{Homi Bhabha National Institute, Training School Complex, Anushakti Nagar, Mumbai 400094, India}


\begin{abstract}
The ability of quantum walks to evolve in a superposition of distinct quantum states has been used as a resource in quantum communication protocols. Under certain settings, the  $k$-cycle discrete-time quantum walks\,(DTQW) are known to recur to its initial state after every $t_r$ steps.   We first present a scheme to optically realize any $k$-cycle DTQW using $J$-plate, orbital angular momentum\,(OAM) sorters, optical switch, and optical delay line. This entangles the polarization and OAM degrees of freedom\,(DoF) of a single photon. Making use of this recurrence phenomena of $k$-cycle DTQW and the entanglement generated during the evolution, we present a new quantum direct communication protocol. The recurrence and entanglement in $k$-cycle walk are effectively used to retrieve and secure the information, respectively, in the proposed protocol. We investigate the security of the protocol against intercept and resend attack. We also quantify the effect of amplitude damping and depolarizing noises on recurrence and mutual information between polarization and OAM DoF of a single photon. Finally, we have indicated an optimal attack strategy by which an Eavesdropper can tamper part of the message without revealing her presence. However, when the quantum communication channel is less noisy, any attempt by the Eavesdropper to tamper the message would end up in exposing her to the receiver.
\end{abstract}


\maketitle

\section{Introduction}

Shor's algorithm for factorization of prime numbers in polynomial time using quantum computers\,\cite{Shor_1997} posed threat to RSA  cryptography for secured communications.  Although quantum secure communications started with BB84 protocol\,\cite{Bennett_2014} to exchange quantum keys, it became important after the discovery of Shor's algorithm. Even though practical implementation of Shor's algorithm is still a long way, the quantum communication schemes have been developing since then\,\cite{gisin2007}. 

The quantum communication protocols are broadly classified into 3 types -- quantum key distribution\,(QKD), quantum secure direct communication\,(QSDC) and quantum dialogue\,(QD). A secret key is initially supplied to the recipient in a QKD protocol, then the message is encoded so that it can be decoded by simply adding the secret key. The first quantum communication protocol that was proposed, BB84, which employed two sets of orthogonal basis states, was a QKD protocol\,\cite{Bennett_2014}. The QKD protocol using non-orthogonal basis was also developed  later on\,\cite{bennett1992}. Subsequently,  entanglement based QKD protocol was also proposed using Bell's state\,\cite{Ekert91}. The second form of communication is QSDC, in which the messages are conveyed directly from one side to the other using quantum channels\,\cite{bostrom2002, srikara2020, long2007,lee2006, deng2004}. The advantage of this method over the previous one is that we don't need a secret key. Various forms of QSDC have been implemented which include one-step\,\cite{sheng2022}, two-step \cite{deng2003} and three-step--three-party\,\cite{chen2018} QSDC. Experimentally, QSDC has been demonstrated with the aid of quantum memory\,\cite{zhang2017}. The third type of protocol is the QD\,\cite{nguyen2004, zhang2006}. In QD, both parties engage in a conversation via quantum channels\,\cite{Saxena_2020, nguyen2004, zhong2005, zhang2006}. Single photon based QD protocol has also been implemented\,\cite{yang2007,luo2014}. The primary advantage of this protocol is that it enables two-way communication in the quantum channel. 

Quantum walks\,(QW) have also been considered as a potential candidate for quantum communication protocols\,\,\cite{yang2018, vlachou2015, vlachou2018, srikara2020}. The evolution of QW is inspired by the classical random walk by embedding the quantum features  like superposition into the dynamics\,\cite{aharonov1993}. The feature of the QW to evolve in superposition of position space has been used in designing various quantum search algorithms\,\cite{santha2008} and portfolio optimization algorithms\,\cite{slate2021}. Recently, the QW has been exploited for various machine learning applications and page ranking applications\,\cite{de2019, prateek2020}, where the QW-based search algorithm is used to find the optimized weights. Due to the recent implementation of QW in two dimensions on a chip, it has now become possible to work with, and design practically implementable quantum communication protocols\,\cite{tang2018}. Implementation of QW has been theoretically proposed and experimentally demonstrated in orbital angular momentum\,(OAM) degree of freedom\,(DoF) of a single photon\,\cite{zhang2007,zhang2010, yasir2022,errico2021}. 

While the position space in one-dimensional discrete-time QW\,(DTQW) is infinite dimensional, we can as well define $k$-cycle DTQW with $k$-dimensional position space. For $k$-cycle DTQW, complete state revival -- the walker returning to the initial position once after every $t_r$ steps -- with particular choice of coin parameters has been shown for different $k$ values such as 2, 3, 4, 5, 6, 8, and 10\,\cite{tregenna2003, chandru2010, dukes2014,bian2017}. This revival can be attributed to quantum recurrence theorem\,\cite{bocchieri57}, which states that any closed quantum system with discrete energy eigenvalues, when it evolves in time, it will repeat itself as accurately as possible after a specific finite time. 

Previously, OAM beam-based communication methods were developed\,\cite{willner2015}. In this paper, we shall discuss a type of QSDC protocol that is practically implementable using QW on OAM states. We first propose an optical setup to perform $k$-cycle DTQW using polarization and OAM DoF of a single photon. Based on the optical setup, we also propose a QSDC protocol. Further, we discuss the possibility of intercept and resend attack. We then demonstrate how recurrence and mutual information between the polarization and the OAM DoF of a single photon are affected with the inclusion of amplitude damping and depolarizing noises. Additionally, we present an optimal attack strategy, where an Eavesdropper can tamper the message without getting detected by the receiver. Nevertheless, any attempt by the Eavesdropper to tamper the message in the presence of less noisy quantum communication channel would just reveal her presence.

The content of rest of this paper is organized as follows. In Section\,\ref{s2}, we describe $k$-cycle DTQW and how it can be realized in polarization and OAM DoF using an optical setup. In Section\,\ref{s3} we explain our direct communication protocol based on $k$-cycle DTQW recurrence. We also propose an implementation scheme in polarization and OAM DoF of a single photon. In Section\,\ref{s4} we discuss the protocol's security and demonstrate how it is resilient against attacks. Finally, in Section\,\ref{s5} we conclude with remarks.

\section{k-cycle DTQW and its realization} \label{s2}

In this Section, we introduce $k$-cycle DTQW and propose an optical setup to realize the same in polarization and OAM DoF of the single photon. DTQW in one dimension essentially consists of two operations\,: coin operation and shift operation. The coin space is spanned by vectors $\{|0 \rangle, |1 \rangle\}$ in the 2-dimensional Hilbert space, whereas the position space is spanned by vectors $\{|x \rangle\}$ in the infinite dimensional Hilbert space, where $x$ can assume any integer values. The coin operator which we shall be interested in our paper is
\begin{align} \label{co}
\hat{C}(\varrho) =  
\begin{bmatrix}
\sqrt{\varrho} & \sqrt{1-\varrho} \\
\sqrt{1-\varrho} & -\sqrt{\varrho}
\end{bmatrix} \otimes \mathds{1}_p,
\end{align} 
where $\mathds{1}_p$ is identity operator in the position space and $0 \leq \varrho \leq 1$\,\cite{dukes2014}. The shift operator is defined as
\begin{align} \label{sf1}
\hat{S} = \sum_x (|0 \rangle \langle 0| \otimes |x-1 \rangle \langle x| + |1 \rangle \langle 1| \otimes |x+1 \rangle \langle x|).    
\end{align}
With these, we can write the evolution of one dimensional DTQW after $t$ steps as
\begin{align} \label{sf2}
|\Psi_t \rangle = [\hat{S} \hat{C}(\varrho)]^t |\Psi_0 \rangle,   
\end{align}
where $|\Psi_0 \rangle$ represents the initial state represented by $[\cos \chi, i\sin \chi]^T \otimes [1,0,\ldots,0]^T_{k \times 1}$.

\begin{figure}[htbp]
    \centering
    \includegraphics[scale=0.53]{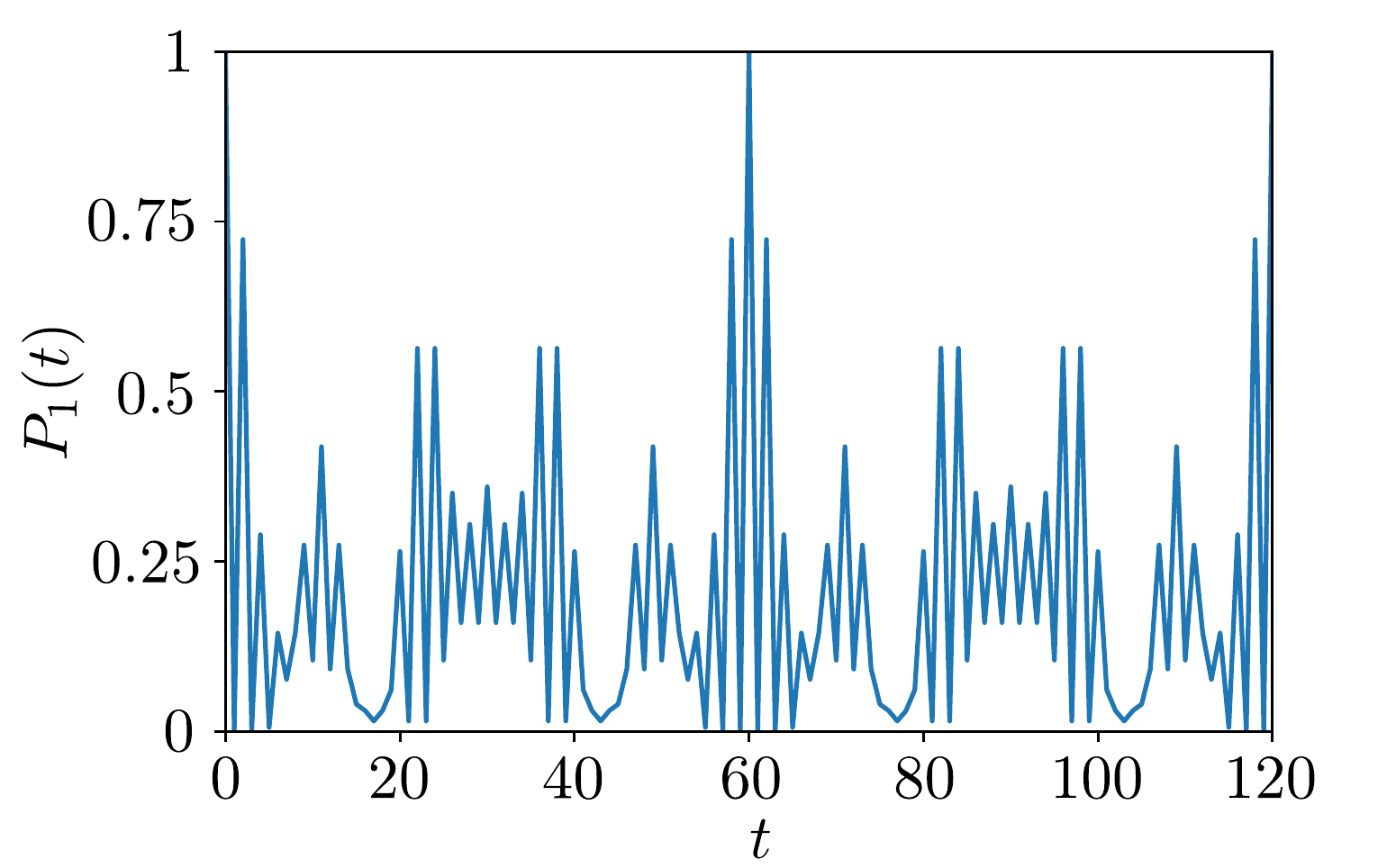}
    \caption{Probability of walker returning to the initial position\,($x=1$), $P_1(t)$, is plotted as a function of steps\,($t$) for a 5-cycle DTQW. The walker returns to the initial position once after every 60 steps.}
    \label{pt1}
\end{figure}

In the case of $k$-cycle DTQW, the position space is a Hilbert space spanned by $k$ vectors, $\{|x \rangle\}$, with $x$ running from $0$ through $k-1$. The shift operator pertaining to this type of DTQW is
\begin{align} \label{sf3}
\hat{S}_k &= \sum_{x=0}^{k-1} (|0 \rangle \langle 0| \otimes |(x-1)\,\,({\rm mod}\,k) \rangle \langle x| \nonumber \\ 
&\,\,\,+ |1 \rangle \langle 1| \otimes |(x+1)\,\,({\rm mod}\,k) \rangle \langle x|).        
\end{align}
After $t$ steps, the evolved state can be written as
\begin{align} \label{sf4}
|\Psi_t \rangle = [\hat{S}_k \hat{C}(\varrho)]^t |\Psi_0 \rangle = \sum_{x=0}^k (a_{x,t} |0 \rangle + b_{x,t} |1 \rangle) \otimes |x \rangle.   
\end{align}
Here, $a_{x,t}$ and $b_{x,t}$ are normalized complex coefficients. It is known that the position space probability distribution recurs completely after $t_r$ steps for specific choices of $k$ and $\rho$. For example, in a $k$-cycle DTQW with $k$ assuming 4, 5 and 6, the position space probability distribution recurs completely after $t_r=20$, 60, and 28 steps, provided $\varrho=(3-\sqrt{5})/8$, $(5-\sqrt{5})/10$, and $2[1-\cos (\pi/7)]/3$, respectively.\,\cite{dukes2014}
This is shown in Fig.\,\ref{pt1}, where the probability of finding the walker at the initial position\,($x=1$), denoted by $P_1(t)$, is plotted against the walk steps. Evidently, the walker returns to the initial position after every 60 steps.  We use the recurrence as a tool to encode and decode the messages for the QSDC protocol discussed below. As we require the same optical configuration for each step of the $k$-cycle DTQW, the benefits of recurrence may be seen as the optimization of the optical setup.

\begin{figure*}
\centering
  \includegraphics[scale=0.28]{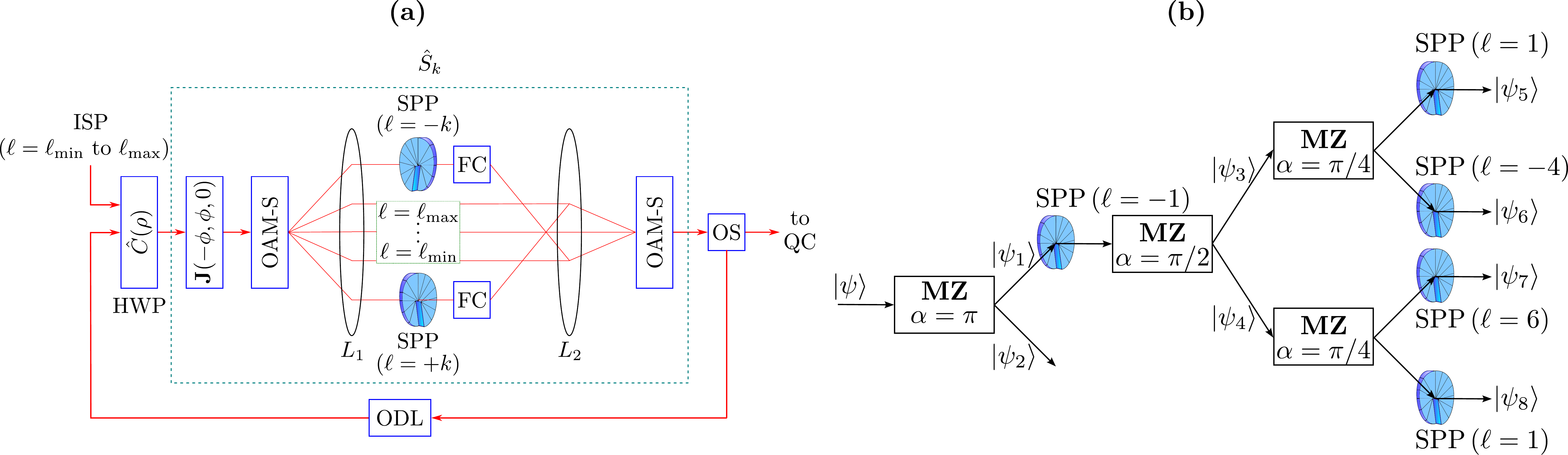}
  \caption{(a) Shown here is the optical realization of $k$-cycle DTQW. An incoming single photon\,(ISP) with OAM value\,($\ell$) between $\ell_{\rm min}$ and $\ell_{\rm max}$ is first sent to the coin operator $\hat{C}(\rho)$\,[see Eq.\,(\ref{co})] -- which is a half-waveplate\,(HWP) rotated through an angle $\beta/2$, with $\beta=\tan^{-1} (\sqrt{(1-\varrho)/\varrho})$. The shift operator $\hat{S}_k$ is realized using a $J$-plate, represented by $\mathbf{J}(-\phi,\phi,0)$\,[see Eq.\,(\ref{j1a})], two OAM sorters\,(OAM-S), thin lenses $L_1$ and $L_2$, two spiral phaseplates\,(SPP) with $\ell=\pm k$, and two fiber couplers\,(FC). Having realized both coin and shift operations, the photon is now sent through an optical switch\,(OS) and optical delay line\,(ODL) before being passed through the HWP once again for the next step. Upon performing the desired number of steps, the photon is finally sent to the quantum channel\,(QC) through the use of OS. (b) Shown here is the Mach-Zehnder\,(MZ) + SPP setup which sorts the incoming OAM modes. When an ISP, represented by $|\psi \rangle$\,[see Eq.\,(\ref{mz1})] passes through a HWP, a $J$-plate, and the MZ + SPP setup, one-step in the 5-cycle DTQW is performed by the ISP. First MZ setup sorts odd and even OAM modes. The SPP with $\ell=-1$ converts all odd OAM modes to even OAM modes. Subsequently, the remaining 3 MZ setups and 4 SPPs with appropriate $\ell$ values ensure the modulo 5-addition of the 5-cycle DTQW. Now the states $|\psi_2 \rangle$, $|\psi_5 \rangle$, $|\psi_6 \rangle$, $|\psi_7 \rangle$, and $|\psi_8 \rangle$ are fiber-coupled together and sent to OS.}
    \label{cqw}
\end{figure*}

Having defined $k$-cycle DTQW, we shall now discuss the optical realization of the same in the remainder of this Section. While the coin operator $\hat{C}(\varrho)$\,[Eq.\,(\ref{co})] is realized using a half-waveplate\,(HWP) rotated through an angle $\beta/2$, with $\beta=\tan^{-1} (\sqrt{(1-\varrho)/\varrho})$\,\cite{hecht2002}, the $k$-cycle DTQW shift operator, $\hat{S}_k$, makes use of a $J$-plate, 2 OAM sorters\,\cite{berkhout2010, mirhosseini2013}, 2 thin lenses, 2 spiral phase plates\,(SPP), and 2 fiber couplers\,(FC). Before explaining our setup in detail, we first briefly introduce $J$-plates.

{\noindent \bf $J$-plates\,:} $J$-plates are used to convert spin angular momentum of the incoming photon into OAM of the same\,\cite{devlin2017,mueller2017}. The Jones matrix of $J$-plate can be given as
\begin{align}
\mathbf{J}(\delta_x,\delta_y,\vartheta) &= R_\vartheta
\begin{bmatrix}
e^{i\delta_x} & 0 \\
0 & e^{i\delta_y}
\end{bmatrix}
R_{-\vartheta}, \label{j1a} \\
{\rm where} \,\,\,
R_\vartheta &= \begin{bmatrix}
\cos \vartheta & -\sin \vartheta \\
\sin \vartheta & \cos \vartheta
\end{bmatrix}. \label{j1b}
\end{align}
At every point in the transverse plane in which the $J$-plate is kept, we can independently provide phase shifts, $e^{i\delta_x}$ and $e^{i\delta_y}$, in $x$- and $y$-directions, respectively. In other words, every point acts as a tiny variable waveplate. Also, every one of these variable waveplates can be independently rotated through an angle $\vartheta$ about the $z$-axis. Suppose we consider a single photon such that its polarization DoF is mapped to the coin Hilbert space and OAM DoF is mapped to the position Hilbert space. We now observe that the shift operator corresponding to the DTQW can readily be realized using the $J$-plate as 
\begin{align} \label{j2}
\mathbf{J}(-\phi,\phi,0) = e^{-i\phi} |H \rangle \langle H| + e^{i\phi} |V \rangle \langle V|,   
\end{align}
where $\phi=\tan^{-1} (y/x)$, with $(x,y)$ being the coordinates of the transverse plane in which the $J$-plate is kept. Also, $|H \rangle$ and $|V \rangle$ are Jones vectors pertaining to horizontal and vertical polarization states, respectively.

Now we shall explain the proposed optical setup, which lets a single photon perform $k$-cycle DTQW for the desired number of steps, as shown in Fig.\,\ref{cqw}\,\tr{(a)}. In the optical setup, an incoming single photon\,(ISP) whose OAM value\,($\ell$) lying between $\ell_{\rm min}$ and $\ell_{\rm max}$ is sent through the HWP rotated through an angle $\tan^{-1} (\sqrt{(1-\varrho)/\varrho})/2$ and the $J$-plate, $\mathbf{J}(-\phi,\phi,0)$. We define $\ell_{\rm min}$ as $\ell_{\rm max}$ as follows\,:
\begin{align} \label{lm}
(\ell_{\rm min}, \ell_{\rm max}) = \left\{ \begin{array}{rcl}
(-\floor{k/2}+1,\floor{k/2}), & \mbox{for} & {\rm even}\,\,k, \\  
(-\floor{k/2},\floor{k/2}), & \mbox{for} & {\rm odd}\,\,k,
\end{array}\right.    
\end{align}
where $\floor{~}$ denotes the floor function. Now the photons with different $\ell$ values are  spatially separated by the first OAM sorter\,(OAM-S). Those photons with $\ell$ values lying between $\ell_{\rm min}$ and $\ell_{\rm max}$ pass through the thin lenses $L_1$ and $L_2$ and are recombined into a single beam through the use of the second OAM-S. Note that the second OAM-S is operated in the reverse direction\,\cite{fickler2014}. Meanwhile, the photons with $\ell=\ell_{\rm max}+1$\,($\ell=\ell_{\rm min}-1$) pass through a SPP with $\ell=-k$\,($\ell=+k$) such that their resultant OAM value is $\ell=\ell_{\rm min}$\,($\ell=\ell_{\rm max}$). By fiber coupling\,(FC) these photons -- whose OAM value is $\ell=\ell_{\rm min}$\,($\ell=\ell_{\rm max}$) -- with those spatially separated by the first OAM-S and $L_1$, we can effectively implement one shift operation for the $k$-cycle QW, $\hat{S}_k$. Along with the coin operator, $\hat{C}(\varrho)$, this just constitutes one step of the $k$-cycle DTQW. Through the use of an optical switch\,(OS)\,\cite{ono2020,guo2022} and an optical delay line\,(ODL), which provides the desired time delay necessary for the single photon, we can once again send the single photon to both coin and step operators such that another step of walk in the $k$-cycle DTQW is performed. After performing $t$-such steps, the OS sends the single photon to a quantum channel\,(QC) for further processing. The OAM-S presented here has been reported to have a sorter efficiency of 92\%\,\cite{mirhosseini2013}. However, there has also been reports of sorter with much higher efficiencies\,\cite{ruffato2018,li2017,wei2020}. 

We can also make use of the Mach-Zehnder\,(MZ) interferometric setup, which has 100\% theoretical efficiency\,\cite{leach2002} to sort the OAM modes of the ISP. We would require a HWP, a $J$-plate, MZ setups to sort the constituent OAM modes, an OS, and an ODL to perform $k$-cycle DTQW. The MZ setup consists of two Dove prisms at one arm such that one of the Dove prisms is rotated through some angle about the $z$-axis. 

In Fig.\,\ref{cqw}\,\tr{(b)} we have shown how an ISP, on passing through the HWP and $J$-plate and before being sent to OS and ODL, will perform one-step in the 5-cycle DTQW using 4 MZ setups and SPPs. In 5-cycle DTQW, the ISP is assumed to be in the state
\begin{align} \label{mz1}
|\psi \rangle = \sum_{\ell=-3}^3 c_\ell |\ell \rangle.    
\end{align}
With the aid of MZ with $\alpha=\pi$, odd and even OAM modes are separated, and the resultant states are $|\psi_1 \rangle=c_{-3}\ket{-3}+c_{-1}\ket{-1}+ c_{1}\ket{1}+c_{3}\ket{3}$ and $|\psi_2 \rangle=c_{-2}\ket{-2} +c_{0}\ket{0} +c_{2}\ket{2}$. Now $|\psi_1 \rangle$ passes through a SPP with $\ell=-1$ and become even OAM modes. When this state is passed through the second MZ with $\alpha=\pi/2$, we obtain $\ket{\psi_3} = c_{-1}\ket{-2} + c_{3}\ket{2}$ and $\ket{\psi_4}=c_{-3}\ket{-4}+c_{1}\ket{0}$. When both $|\psi_3 \rangle$ and $|\psi_4 \rangle$ are sent through two different MZ setups with $\alpha=\pi/4$, the constituent OAM modes will be sorted. Finally, when these OAM modes pass through the SPPs with appropriate $\ell$ values -- which ensure modulo 5 addition -- as shown in the Fig.\,\ref{cqw}\,\tr{(b)}, the resultant sorted OAM modes are $|\psi_5 \rangle=c_{-1} |-1 \rangle$, $|\psi_6 \rangle=c_3 |-2 \rangle$, $|\psi_7 \rangle=c_{-3}|2 \rangle$, and $|\psi_8 \rangle=c_1|1 \rangle$, respectively. Now $|\psi_2 \rangle$, $|\psi_5 \rangle$, $|\psi_6 \rangle$, $|\psi_7 \rangle$, and $|\psi_8 \rangle$ are fiber coupled together and sent to OS. We assume that the coherence length of the laser used in the optical setup is large enough such that the fiber coupling of these OAM modes are possible.

We can conclude this section with an interesting feature of our experimental setup. Since the $k$-cycle DTQW revives completely after every $t_r$ steps, performing both $t$-steps as well as $(t+t')$-steps, where $t'$ is just a multiple of $t_r$, will yield the same probability distribution in the position space.

\begin{figure*}
\centering
  \includegraphics[scale = 0.35]{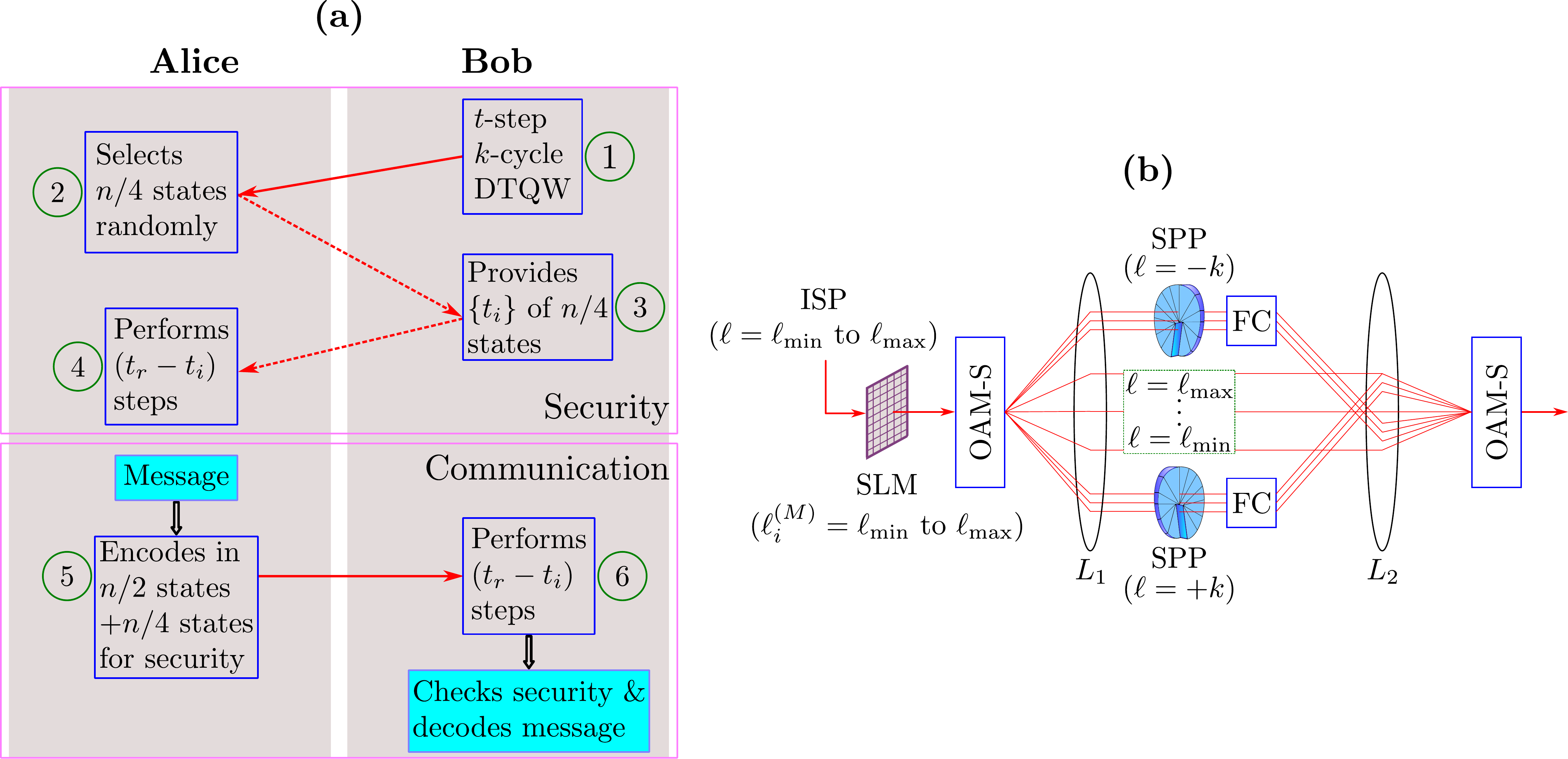}
  \caption{(a) Shown here is the schematic outline of the communication protocol. Here, bold arrows denote quantum channel, and broken arrows denote classical channels. (b) Shown here is the optical setup which adds\,(mod $k$ addition) an OAM value between $\ell_{\rm min}$ and $\ell_{\rm max}$ to the incoming single photon\,(ISP) through the use of a spatial light modulator\,(SLM).} 
    \label{pcl}
\end{figure*}

\section{Communication protocol} \label{s3}

In this Section, we propose a direct communication protocol using $k$-cycle DTQW and an optical setup which encodes message in OAM DoF of the single photon. As mentioned in the introduction, direct communication protocols do not require the sharing of a secret key. As opposed to this, the receiver directly decodes the message once it has been encoded in a state. For instance, in QSDC, the message is typically encoded by performing a unitary operation on a given initial state. The decoding is done by performing another unitary operation that restores the initial state. Our protocol consists of six steps as schematically outlined in Fig.\,\ref{pcl}\,\tr{(a)}, and we enumerate them below.

\begin{enumerate}
\item Before beginning the algorithm, both Alice and Bob {\it a priori} decide the OAM value $\ell_i$ of each of $n$ single photons, where $\ell_i$ can be randomly chosen between $\ell_{\rm min}$ and $\ell_{\rm max}$. First, Bob generates $n$ single photons\,(each with an OAM of $\ell_i \hbar$ per photon) and performs $k$-cycle DTQW on each one of them using the optical setup shown in Fig.\,\ref{cqw} for $t_1$, $\ldots$, $t_n$ steps, respectively, where each $t_i=\{2,3,\ldots, t_r\}$. 
\item These $n$ single photons are now sent to Alice through a quantum channel, and she selects $n/4$ single photons randomly. 
\item Alice requests Bob through a classical channel to provide $t_i$ of each of these $n/4$ states, and Bob sends them through a classical channel. Note that the $t_i$ values are communicated only after Alice confirms that she has received the states. She immediately closes the quantum channel for any other recipient. 
\item Alice now performs remaining $(t_r-t_i)$ steps on each of these $n/4$ single photons and measures OAM of them. If every single photon possesses an OAM of $\ell_i \hbar$ per photon, then there was no eavesdropping and they continue with their protocol; otherwise, they abort and begin all over again. These four steps constitute the {\it security} part of our communication protocol.
\item Alice then randomly chooses $n/2$ out of $3n/4$ states and encodes the intended message in them. Here, the remaining $n/4$ dummy states are used for security purposes as detailed in the next step. It can be remarked that there is no specific reason to choose $n/4$ dummy states out of $3n/4$ states for security purposes. It is just that more number of dummy states would ensure that the security of the protocol is not compromised.
\item Finally, Bob performs remaining $(t_r-t_i)$ steps on each of these single photons to decode the message sent by Alice. By receiving the coordinates of the dummy states from Alice, Bob now compares whether the OAM of these dummy states match with the ones he initially sent to Alice. With this, Bob can ensure eavesdropping. Therefore, the last two steps of our protocol constitute the {\it communication} part.
\end{enumerate}

{We now have the following remark. Because Alice {\it a priori} knows the distance between her and Bob, she can ensure that she opens and closes the quantum channel within the {\it time-window}. This readily guarantees that the states sent by Eve won't interfere with those sent by Bob during the post-processing stage.}

We now explain how Alice encodes message using optical setup shown in Fig.\,\ref{pcl}\,\tr{(b)}. We assume that the ISP possesses $\ell$ between $\ell_{\rm min}$ and $\ell_{\rm max}$, and we wish to encode message $\ell_{i}^{(M)}$ on the $i$-th single photon, where $\ell_{i}^{(M)}$ assumes integer values between $\ell_{\rm min}$ and $\ell_{\rm max}$. The addition of $\ell_{i}^{(M)}$ to the single photon can be provided using an SLM\,\cite{saleh2007}. When $\ell+\ell_{i}^{(M)}$ falls between $\ell_{\rm min}$ and $\ell_{\rm max}$, the resultant OAM is not affected by the optical setup followed by the SLM. We note that the message encoding can also be performed using the MZI setup\,(see Fig.\,\ref{cqw}\,\tr{(b)}) as well instead of the OAM-S. Otherwise, the two SPPs and the two FCs will suitably add or subtract an OAM value $k$ and couple the resultant OAM modes with those falling between $\ell_{\rm min}$ and $\ell_{\rm max}$. Thus, the optical setup following the SLM ensures that the addition $\ell+\ell_{i}^{(M)}$ is indeed a modulo-$k$ addition. Since Bob knows $\ell_i$ as well as $\ell_i+\ell_{i}^{(M)}\,({\rm mod}\,k)$ for each of $3n/4$ single photons, he can easily deduce $\ell_{i}^{(M)}$. In Appendix\,\ref{me} we show that the message encoding operation by Alice can be performed after any number of steps, $t_i$, with $2 \leq t_i \leq t_r$, and consequently, Bob will unambiguously decode the message.
 
{If we are employing $k$-cycle DTQW, we must first transform the messages to strings of the basis $k$ before encoding them into the OAM states. We know that any message may be converted into a string of $k$ bases by creating an one-one and onto function. For example, if we use $k=8$-cycle DTQW and we wish to encode the message ``It is sunny today'', we can convert it to a string using ASCII to octal converter. Now the message will read ``111 164 040 151 163 040 163 165 156 156 171 040 164 157 144 141 171''. Likewise, we can use ASCII to any $k$-base converter to encode the message in OAM state.}

\section{Security of the protocol} \label{s4}

In this Section, we first discuss intercept and resend attack and calculate how much information can be eavesdropped by a third party for various $k$-cycle DTQW. We then explore how various levels of noises in the polarization DoF of the single photon affect the recurrence and the entanglement between the polarization and the OAM DoF of the single photon. In general, quantum communication protocols take into account attacks like intercept and resend\,\cite{chang2015}, man in the middle\,\cite{peev2004}, trojan horse\,\cite{deng2005}, and denial of service\,\cite{boche2021}. Some of these attacks are discussed for QKD schemes too\,\cite{scarani2009,sun2022}. The quantum communication protocols are thought to be most vulnerable to intercept and resend attacks, which include circumstances where security can be easily breached. We first begin with the intercept and resend attack.

\subsection{Intercept and resend attack during sending}

\begin{figure}[htbp]
    \centering
    \includegraphics[scale=0.53]{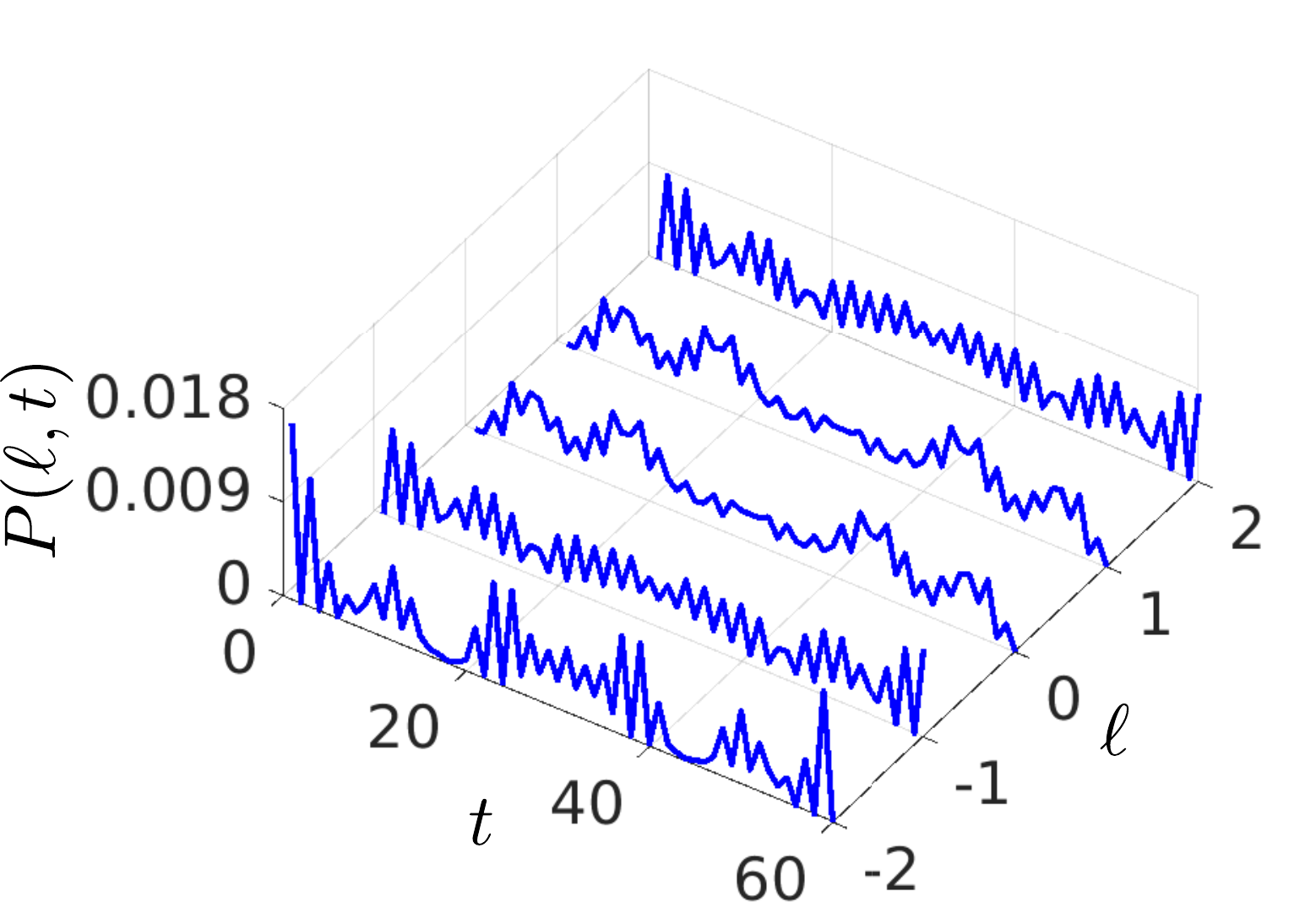}
    \caption{Shown here is the joint probability distribution, $P(\ell,t)$, of finding the walker at position $\ell$ after $t$ steps corresponding to a 5-cycle DTQW with coin parameter $\varrho=(5-\sqrt{5})/10$. The probability of finding the walker at a given $t$ over all $\ell$ values is 1. Because we want the sum over all possible $\ell$ and $t$ for $P(\ell,t)$ to be 1\,[see Eq.\,(\ref{pb})], we have divided the probability distribution by $(t_r-1)$.}
    \label{pt}
\end{figure}

In this attack, Eve intercepts the states sent by Bob and resend them to Alice. Because she does not know $t_i$ of each of the $n$-photons, she guesses the same, which we denote as $t_i^{(E)}$. Now she performs remaining $(t_r-t_i^{(E)})$ steps and measures the OAM value of $i$-th photon as $\ell_i^{(E)}$. She then prepares a single photon with OAM value $\ell_i^{(E)}$, performs $t_i^{(E)}$-step $k$-cycle DTQW, and sends it to Alice. Since $t_i=\{2,3,\ldots, t_r\}$, the probability of Eve guessing $t_i$ of each photon is $P(t_i^{(E)})=1/(t_r-1)$. For brevity, we shall just denote $P(t_i^{(E)})$ as $P(t_i)$. Further, the probability of finding the walker at any given $t$ is 1, as the walker would be at any one of the position spaces. In Fig.\,\ref{pt} we have shown the probability of finding the walker\,(performing the 5-cycle DTQW) at various positions for $t_r=60$ steps. Because this is a joint probability distribution in $\ell$ and $t$, we have just divided the distribution by $1/(t_r-1)$ so that
\begin{align} \label{pb}
\sum_{\ell=\ell_{\rm min}}^{\ell_{\rm max}} \sum_{t=2}^{t_r} P(\ell,t) = 1.    
\end{align}
Now $P(\ell)$, namely, the probability of Eve guessing the OAM value of the single photon, can be easily read off by summing over $t$ axis, i.e.,

\begin{align} \label{pl}
P(\ell) &= \sum_{t=2}^{t_r} P(\ell,t).
\end{align}
Therefore, when Eve guesses the step number $t$, the mutual information\,\cite{nielsen2010} gained by her about the OAM value $\ell$ is

\begin{figure*}
    \centering
\includegraphics[scale=0.55]{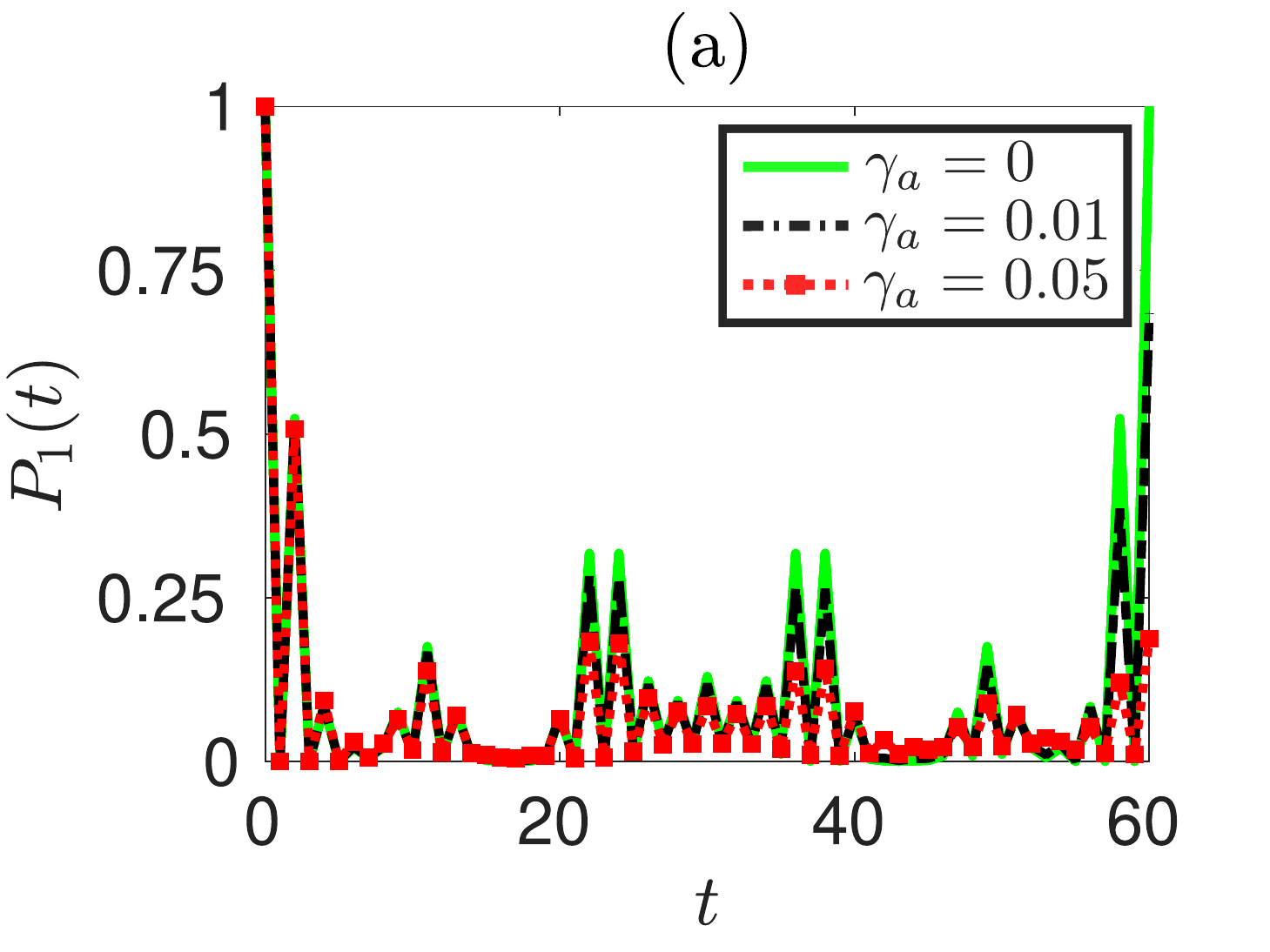}~~
\includegraphics[scale=0.55]{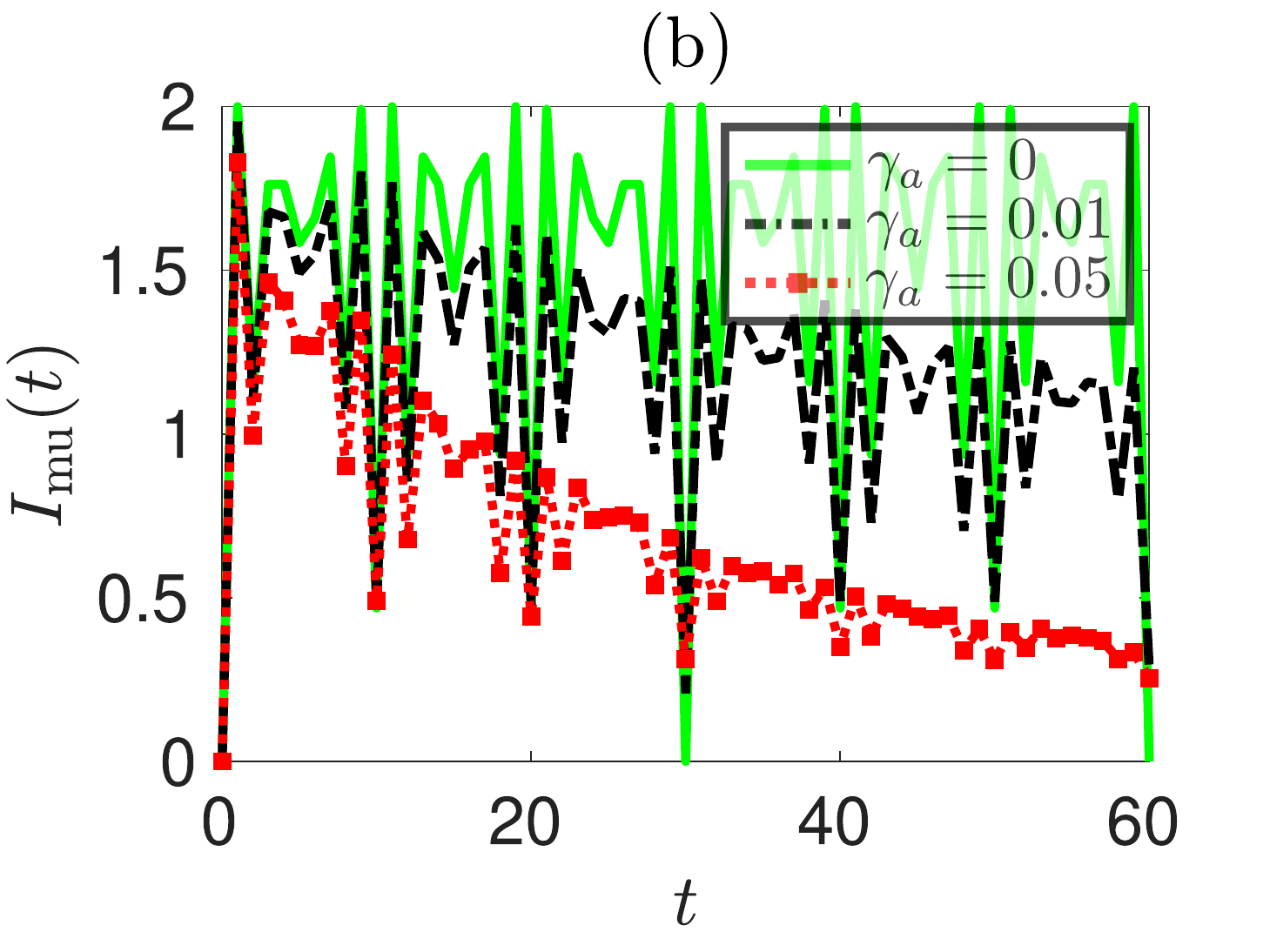}\\
\includegraphics[scale=0.55]{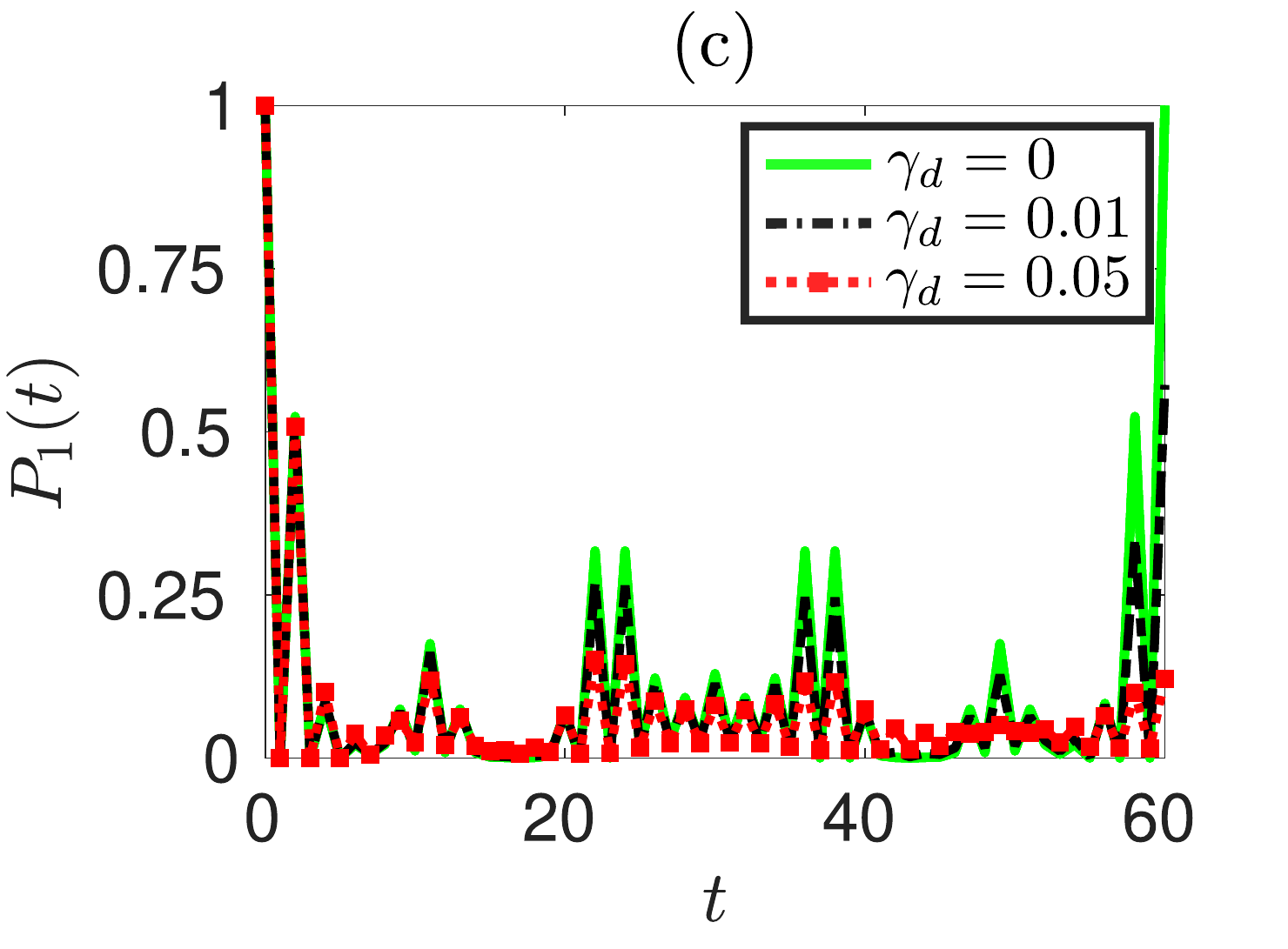}~~
\includegraphics[scale=0.55]{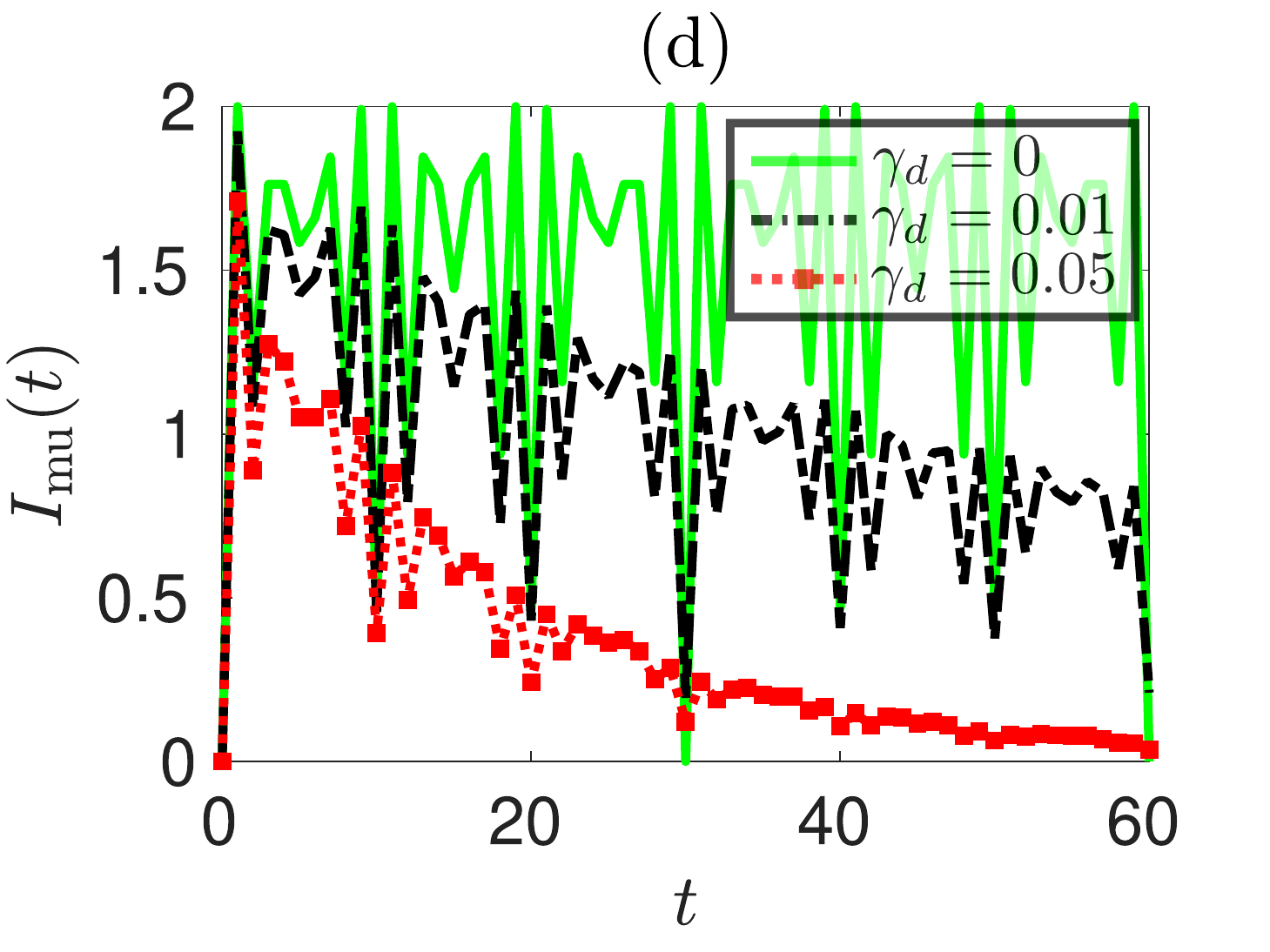}
    \caption{(a) Shown here is the probability of the $5$-cycle walker returning to the position $x=1$, $P_1(t)$, after every step $t$ for various levels of amplitude damping noise parameter $\gamma_a$\,[see Eq.\,(\ref{ns1})]. (b) Shows how the mutual information, $I_{\rm mu}(t)$\,[see Eq.\,(\ref{imu})], varies as a function of $t$ for the same amount of amplitude damping noise parameters. In (c) and (d) we have plotted $P_1(t)$ and $I_{\rm mu}(t)$ for various levels of depolarizing noise parameter $\gamma_d$\,[see Eq.\,(\ref{ns2})], respectively.}
    \label{pmu}
\end{figure*}

\begin{align} \label{im}
I_{t,\ell,E} = \sum_{t=2}^{t_r} \sum_{\ell=\ell_{\rm min}}^{\ell_{\rm max}} P(\ell,t)\log_2 \left[ \frac{P(\ell,t)}{P(\ell)P(t)} \right].
\end{align}
The mutual information, $I_{t, \ell,E}$, quantifies how much information we can gain about one random variable\,(say, $\ell$) by guessing the other random variable\,(say, $t$). As is known, small value of $I_{t, \ell,E}$ indicates that Eve cannot know more about the OAM value, $\ell$, by guessing the number of steps, $t_i$, and vice versa. In Table\,\ref{table} we have listed out the mutual information, $I_{t,\ell,E}$, for some known $k$-cycle DTQW.

\begin{table}[htbp]
\centering
\begin{tabular}{ccccc}
Cycles\,($k$) & $\varrho$ & $t_r$ & $P(\ell=\ell_{\rm min})$& $I_{t, \ell,E}$ \\
\hline
\hline
3 & 2/3 & 8 & 0.238095 & 0.174429 \\
4 & $(3-\sqrt{5})/8$ & 20 & 0.210526 & 1.175981 \\
5 & $(5-\sqrt{5})/10$ & 60 & 0.186441 & 0.358934 \\
6 & $2[1-\cos (\pi/7)]/3$ & 28 & 0.220722 & 1.218855\\
8 & 1/2 & 24 & 0.108696 & 1.281487 \\
10 & $(5-\sqrt{5})/10$ & 60 & 0.131488 & 1.317215 \\
\hline
\hline
\end{tabular}
\caption{In this table, we have enumerated probability of measuring the OAM value $\ell=\ell_{\rm min}$, $P(\ell=\ell_{\rm min})$\,[see Eq.\,(\ref{pl})], and mutual information gained by Eve about the OAM value $\ell$ on guessing $t$, $I_{t, \ell,E}$\,[see Eq.\,(\ref{im})], for various $k$-cycle DTQW. The corresponding coin parameter $\varrho$\,[Eq.\,(\ref{co})] and the number of steps\,($t_r$) after which the given $k$-cycle DTQW revives completely are also listed out.}
\label{table}
\end{table}

\begin{figure}[htbp]
\centering
\includegraphics[scale=0.45]{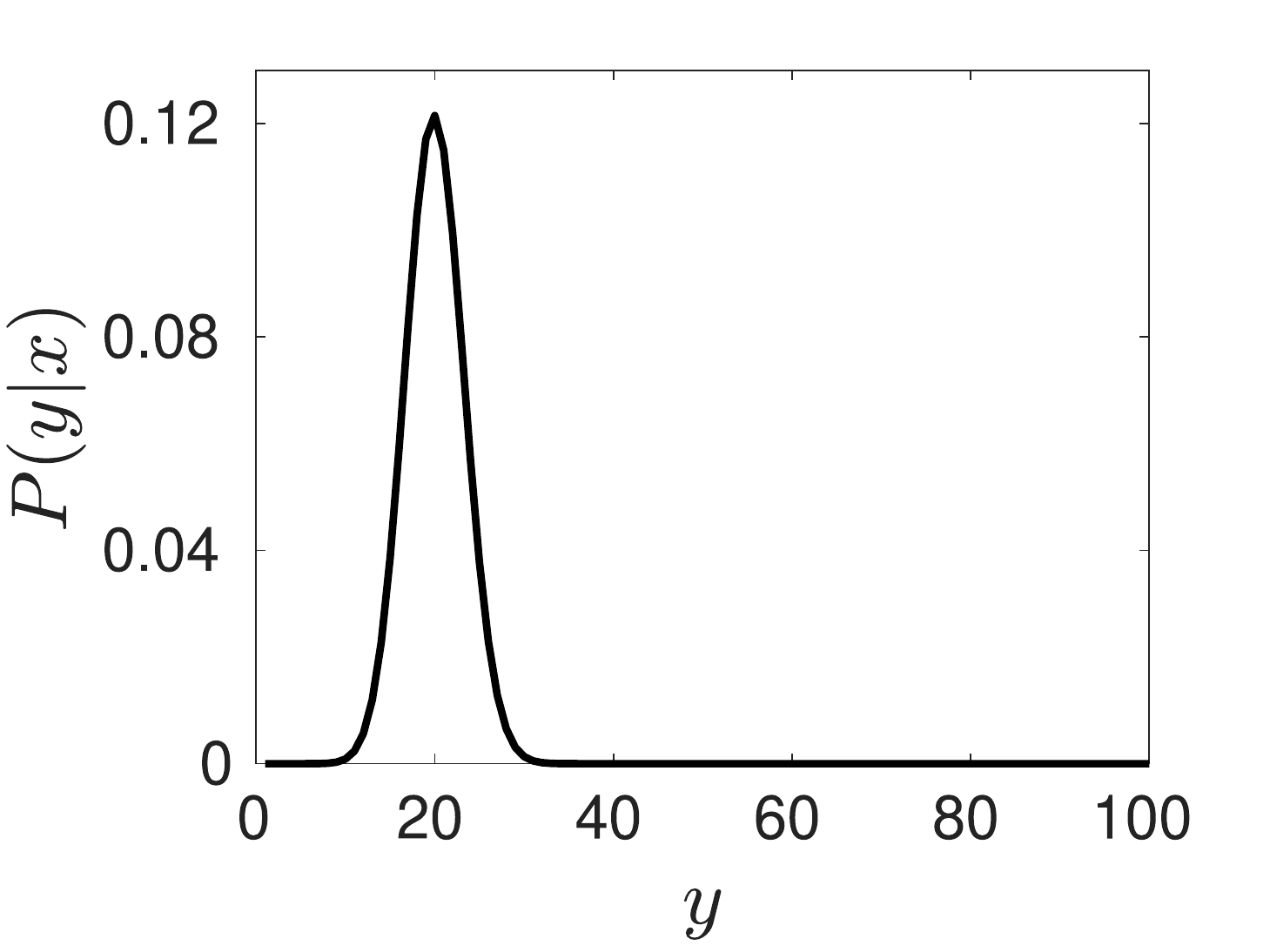}
\caption{This graph plots $P(y|x)$ versus $y$\,[see Eq.\,(\ref{pyx})] for the choices $N=100$ and $x=60$. $P(y|x)$ looks like a Gaussian distribution centered at $y=20$.}
\label{sty}
\end{figure}

\subsection{Mutual information between coin and OAM DoF}

One convenient measure that captures entanglement between the polarization and OAM DoF of a single photon is mutual information, $I_{\rm mu}$, and is defined as\,\cite{nielsen2010},
\begin{align} \label{imu}
I_{\rm mu}(t) = S(\rho^{\rm (pol)}_t) + S(\rho^{\rm (OAM)}_t) - S(\rho_t).
\end{align}
Here, $\rho_t=|\Psi_t \rangle \langle \Psi_t|$\,[see Eq.\,(\ref{sf4})], $\rho^{\rm (pol)}_t$ and $\rho^{\rm (OAM)}_t$ are reduced density matrices obtained after tracing out OAM and polarization DoF of $\rho_t$, respectively, and $S(\cdot)$ computes the von Neumann entropy of the given density matrix. It can be noted that zero mutual information indicates that there is no quantum correlation present between both polarization and OAM DoF of a single photon. For instance, in the case of 5-cycle DTQW, we observe that $I_{\rm mu}(t) \rightarrow 0$ when $t=30$\,(see Fig.\,\ref{pmu}\,\tr{(b)}). We confirmed zero quantum correlation at $t=30$ by computing the entanglement negativity\,\cite{vidal2002} between the polarization and OAM DoF of the single photon. Indeed, at $t=30$ steps, the separable single photon state is given by $([1,i]^T/\sqrt{2}) \otimes ([-3,2,2,2,2]^T/5)$.

\subsection{Amplitude damping and depolarizing noises}

Any signal transmitted between two parties in a communication protocol is susceptible to noise. Two well-known noise models that are used to study how polarization DoF of a single photon is affected are amplitude damping noise and depolarizing noise\,\cite{nielsen2010}. In the former one, the reduced density matrix $\rho_t^{\rm (pol)}$, i.e., density matrix obtained by partial tracing out the OAM DoF in $\rho_t=|\Psi_t \rangle \langle \Psi_t|$\,[see Eq.\,(\ref{sf4})], is 
\begin{align} \label{ns1}
\rho_t^{\rm (pol)} \mapsto E_0 \rho_t^{\rm (pol)} E_0^\dagger + E_1 \rho_t^{\rm (pol)} E_1^\dagger, 
\end{align}
where $E_0 = \begin{bmatrix}
        1 & 0\\
        0 & \sqrt{1-\gamma_a}
    \end{bmatrix}$ and $E_1 = \begin{bmatrix}
        0 & \sqrt{\gamma_a}\\
        0 & 0        
    \end{bmatrix}$, and $0 \leq \gamma_a \leq 1$. In the latter one, $\rho_t^{\rm (pol)}$ is mapped to
\begin{align} \label{ns2}
\rho_t^{\rm (pol)} \mapsto (\gamma_d/2) \mathds{1}_2 + (1-\gamma_d) \rho_t^{\rm (pol)},
\end{align}
with $0 \leq \gamma_d \leq 1$. Figs.\,\ref{pmu}\,\tr{(a)} and \ref{pmu}\,\tr{(c)} show how the probability of the 5-cycle walker returning to the initial position\,($x=1$; see Fig.\,\ref{pt1}) is affected with the addition of varying depolarizing and amplitude damping noise levels. When less amount of noise is inserted\,(i.e., $\gamma_a=\gamma_d=0.01$), the probability of walker returning to the initial position $x=1$ is greater than 0.5 after 60 steps for both amplitude damping and depolarizing noise cases. On the other hand, the probability of the same drops almost to zero after 60 steps in the presence of more noise\,($\gamma_a= \gamma_d=0.5$) for both amplitude damping and depolarizing noise cases. 

Figs.\,\ref{pmu}\,\tr{(b)} and \ref{pmu}\,\tr{(d)} plot how the mutual information between the polarization and OAM DoF of the single photon, $I_{\rm mu}(t)$, vary as a function of walker steps when different noise levels are injected. Due to the presence of noise, we observe that $I_{\rm mu}(30) \neq 0$. When the inserted noise is high\,($\gamma_a=\gamma_d=0.5$), the mutual information $\rightarrow 0$ for both depolarizing and amplitude noise cases, and consequently, the entanglement between the polarization and OAM DoF is lost.

\subsection{Optimal attack strategy}

Now we devise a strategy by which Eve can intercept the message sent by Alice such that Bob is unable to detect Eve's presence. Before further analysis, we just estimate the probability by which Eve can intercept the message sent by Alice and still not getting detected during when Bob checks the states intended for security purposes. This is possible when Eve wishes to intercept $n/2$ states alone in which Alice has encoded the message\,(see Fig.\,(\ref{pcl}\,\tr{(a)})) and replaces it with her own message. In this case, Eve has to correctly guess both $n/2$ out of $3n/4$ message states as well as the $t_i$'s of each of those message states. The probability by which Eve successfully decrypts the original message sent by Alice completely is then given by
\begin{align} \label{scs}
P({\rm Eve\,\,decrypts}) = \frac{1}{^{3N} C_{2N}} \times \frac{1}{(t_r-1)^{2N}},
\end{align}
where $N=n/4$. Thus, it is highly unlikely that Eve can decrypt the entire message for reasonably large values of $n$.

Because the above scenario is unlikely, we consider the following situation in which Eve can tamper part of the message sent by Alice without being detected by Bob. We further assume that the quantum channel used for communication by Alice and Bob suffers from noise. For brevity, we model this as an amplitude damping noise with $\gamma_a= 0.0007$. The probability of 5-cycle walker returning to the initial position can be found out to be 0.97 for $\gamma_a=0.0007$. Therefore, on an average, we can expect about 97 out of 100 states to recur completely in the presence of this noise. Now suppose Eve decides to randomly select $x$ out of $3N$ states and modifies those states. Since $x$ is chosen randomly, we may expect roughly $N/(3N)=1/3$ states to be selected by Eve as dummy states. To be precise, the probability of $y$ states among those $x$ states being dummy states is
\begin{align} \label{pyx}
P(y|x) = \frac{ ^{N}C_{y} \times ^{2N}C_{x-y}}{ ^{3N}C_{x}}.
\end{align}
In Fig.\,\ref{sty} we have plotted $P(y|x)$ for $x=60$ and $N=100$. It is clear that $P(y|x)$ resembles a Gaussian distribution centered about $y=x/3=20$. This obviously implies that Eve should essentially choose less number of states to tamper with so that Bob does not detect her presence. However, given that Eve chooses to tamper just $x=3$ states, Bob would detect about $y=1$ dummy state and think that 3 message states have been tampered due to noise. Hence, in the noiseless case, any attempt by Eve to tamper the message will reveal her presence to Bob.

{We now demonstrate how noise can alter the intended message and how it can be overcome. If we assume that 100 states are sent in the presence of the noise level described above, 97 of them would be recurring completely, while the remaining 3 would not. However, by adopting the \textit{majority voting}\,\cite{nielsen2010} scheme, we could avoid this. Evidently, as the message is transmitted more frequently, the risk that the same part of the message will be obstructed diminishes. On receiving the same message several times, Bob would employ the  
majority voting scheme to correctly decode the message.}

\section{Concluding remarks} \label{s5}

We have proposed a quantum direct communication protocol making use of the recurrence phenomenon in a $k$-cycle DTQW. We have analyzed our protocol against intercept and resend attack as well as shown how our protocol is robust against few optimal attack strategies. The realization of $k$-cycle DTQW makes use of an optical setup, which implements the QW in the polarization and OAM DoF corresponding to a single photon. Here, the single photon, which can be in a superposition of $k$ OAM states, just requires one quantum channel for transmission. Since a $k$-cycle DTQW can be realized in polarization and path DoF of a single photon too\,\cite{bian2017}, the proposed protocol can also be implemented using these two DoF. However, this realization scheme will instead require $k$ quantum channels for transmitting the same amount of information. Making use of this optical scheme for realizing recurrence of probability distribution in the $k$-cycle DTQW's, we have devised our protocol to send information in the OAM DoF of a single photon. A new optical setup has also been proposed to encode information in the OAM DoF of a single photon. We note that 1D DTQW -- in polarization and OAM DoF -- has been experimentally demonstrated up to 14 steps\,\cite{errico2021}. Therefore, $k$-cycle DTQW with fewer number of recurrence steps\,(for example, $k=3$ and $t_r=8$) can be experimentally implemented. Consequently, our QSDC protocol can readily be experimentally demonstrated using our proposed optical setup. These kind of protocols can also be applied to  various other applications including quantum e-commerce\,\cite{thapliyal2019}, quantum voting\,\cite{thapliyal2017, xue2017} and quantum internet\,\cite{kimble2008, wehner2018}.


\appendix

\section{Message encoding operation and choice of $t_i$} \label{me}

In Section\,\ref{s3} we mentioned that Bob can perform $t_i$ number of steps in a $k$-cycle DTQW and send it to Alice. Alice would then encode the message on the single photon and send it back to Bob. Finally, Bob would perform the remaining $(t_r-t_i)$ steps on the single photon and decode the message. In this Appendix we prove that how the message encoded by Alice on a single photon does not explicitly depend on the choice of $t_i$, where $t_i$ can be chosen randomly from the set $\{2, \ldots, t_r\}$. Mathematically, this message encoding operation can be cast as
\begin{align} \label{me1}
\hat{T}_{\ell_i}^{(M)} = \mathds{1}_2 \otimes \sum_{\ell=0}^{k-1} |(\ell+\ell_{i}^{(M)})\,\,({\rm mod}\,k) \rangle \langle \ell|,
\end{align}
where $\mathds{1}_2$ is 2-dimensional identity operator acting on the polarization space and $0 \leq \ell_{i}^{(M)} \leq (k-1)$. We now show that $\hat{T}_{\ell_i}^{(M)}$ commutes with the $k$-cycle DTQW evolution operator mentioned in Eq.\,(\ref{sf4}). This means that the message encoding operation can be performed after any number of steps, $t_i$, and Bob will be able to decode $\ell_{i}^{(M)}$. It is obvious that $\hat{T}_{\ell_i}^{(M)}$ commutes with the coin operator, $\hat{C}(\varrho)$\,[Eq.\,(\ref{co})], because the former only acts on the OAM space. To show that $\hat{T}_{\ell_i}^{(M)}$ commutes with the shift operator $\hat{S}_k$\,[Eq.\,(\ref{sf3})], we find, after a bit of algebra, that
\begin{align} \label{me2}
\hat{T}_{\ell_i}^{(M)} \hat{S}_k &= \sum_{\ell=0}^{k-1} [|0 \rangle \langle 0| \otimes |(\ell+\ell_{i}^{(M)}-1)\,\,({\rm mod}\,k) \rangle \nonumber \\
&\,\,\, + |1 \rangle \langle 1| \otimes |(\ell+\ell_{i}^{(M)}+1)\,\,({\rm mod}\,k) \rangle] \nonumber \\
&= \hat{S}_k \hat{T}_{\ell_i}^{(M)}. 
\end{align}
Thus, we see that both $\hat{T}_{\ell_i}^{(M)}$ and $\hat{S}_k$ commute. Suppose Bob performs $t_i$ number of steps on a single photon, represented by the initial state $|\Psi_0 \rangle$, and sends it to Alice. Alice then encodes the message on the single photon and sends the same back to Bob. Bob now performs $(t_r-t_i)$ steps on the single photon which he received. With these and exploiting the fact that $\hat{T}_{\ell_i}^{(M)}$ commutes with $\hat{S}_k$, the resulting state, $|\Psi_{t_r} \rangle$, can be written as
\begin{align} \label{me3}
|\Psi_{t_r} \rangle = [\hat{S}_k \hat{C}(\varrho)]^{t_r-t_i} \hat{T}_{\ell_i}^{(M)} [\hat{S}_k \hat{C}(\varrho)]^{t_i} |\Psi_0 \rangle = \hat{T}_{\ell_i}^{(M)} |\Psi_0 \rangle.
\end{align}
Hence, we conclude that the message encoding by Alice can be performed after any number of steps in a $k$-cycle DTQW so that Bob can unambiguously decode the message sent by Alice.

\begin{acknowledgments}
We acknowledge the support from the Office of Principal Scientific Advisor to Government of India, project no. Prn.SA/QSim/2020 and Interdisciplinary Cyber Physical Systems (ICPS) program of the Department of Science and Technology, India, Grant No.: DST/ICPS/QuST/Theme-1/2019/1 for the support.
\end{acknowledgments}



%

\end{document}